\documentstyle[epsfig]{aipproc}
\begin{document}
\title{Quark Structure of Light Mesons}

\author{Bernard Metsch}
\address{Institut f\"ur Theoretische Kernphysik\\
Universit\"at Bonn\\
Nu{\ss}allee 14-16, D53115 Bonn, Germany}

\maketitle

\begin{abstract}
  On the basis of the Bethe--Salpeter Equation we developed a
  covariant constituent quark model, with confinement implemented by a
  linear potential and an instanton induced interaction explaining
  mass splittings and mixing of pseudoscalar mesons. In addition this
  interaction yields a scalar (essentially) flavour singlet state at
  approximately 1 GeV, considerably lower in mass than the
  corresponding octet states calculated around 1.4 GeV. The validity
  of the present approach was checked through various electroweak
  observables. The puzzling properties of scalar mesons is briefly
  discussed.
\end{abstract}

A glance at the experimental meson spectrum shows two general
features, each with a conspicuous exception: Firstly, states belonging
to an orbital angular momentum multiplet exhibit only small spin-orbit
splittings: e.g. $f_1(1285)$ and $f_2(1270)$, $a_3(2050)$ and
$a_4(2040)$, $K_2(1770)$ and $K_3(1780)$. Exceptional are the low
positions of $f_0(980)$ and $a_0(980)$. Secondly, every isovector
state has an isoscalar partner almost degenerate in mass: e.g.
$\rho(770)$ and $\omega(782)$, $h_1(1170)$ and $b_1(1235)$ up to
$a_6(2450)$ and $f_6(2510)$, reflecting the fact that the inter-quark
forces are flavor symmetric. An exception is of course the huge
splitting of the pseudoscalar mesons $\pi-\eta-\eta'$. Any hadron
model should account for this phenomenology, especially in view of the
identification of exotics, like hybrids, dimesonic states or
glueballs.

The most successful model in this respect is certainly the
non-relativistic constituent quark model, where it is assumed that
excitations of hadrons are effectively described in terms of
constituent quarks interacting through potentials. In a particular
version, where confinement was modeled by a linearly rising
string-like potential and the widely used Fermi-Breit interaction
based on One-Gluon-Exchange was substituted by an instanton induced
force~\cite{Bla90} one can indeed arrive at a satisfactory description
of both meson and baryon spectra. Here, the bulk of states is
determined by the confinement potential alone (thus avoiding large
spin-orbit splittings) and the instanton induced force selectively
acts on pseudoscalar states and accounts for mixing and splitting of
the isoscalars. However, this approach can be criticized because
binding energies compared to the constituent quark masses can be very
large and the Schr\"odinger wave functions are incorrect at large
energies or momentum transfers.  Moreover, although the dilepton
widths of vector mesons can be accounted for in the non-relativistic
approach, see Table~\ref{caption:table1} weak decay constants are too
large by an order of magnitude and the $\gamma\gamma$-decay results
are even beyond discussion. However, a drastically improved description
is found in the relativistically covariant quark model we will now
briefly discuss.

\begin{table}
        \caption{Electro-weak meson decays.}\label{caption:table1}
        \begin{tabular}{lccclccc}
        \multicolumn{1}{c}{Decay}& Exp.~\cite{PDG96}&
        \multicolumn{1}{c}{RQM} &
        \multicolumn{1}{c}{NRQM}&
\multicolumn{1}{c}{Decay}& Exp.~\cite{PDG96}&
        \multicolumn{1}{c}{RQM} &
        \multicolumn{1}{c}{NRQM} \\
        \tableline
        $f_\pi$[MeV] & 131.7 $\pm$ 0.2 & 130 & 1440 &
        $f_K  $[MeV] & 160.6 $\pm$ 1.4 & 180 &  730 \\
        $\Gamma_{\pi^0 \to \gamma\gamma}$[eV] & 7.8 $\pm$ 0.5 & 7.6 &
        30000 &
        $\Gamma_{\rho \to e^+ e^-}$[keV]        & 6.8 $\pm$ 0.3 & 6.8 & 
        8.95 \\
        $\Gamma_{\eta \to \gamma\gamma}$[eV] & 460 $\pm$ 5 & 440 &
        18500 &
        $\Gamma_{\omega \to e^+ e^-}$[keV]        & 0.60 $\pm$ 0.02 &
        0.73 &  0.96 \\
        $\Gamma_{\eta' \to \gamma\gamma}$[eV] & 4510 $\pm$ 260 & 2900 &
        750 &
        $\Gamma_{\phi \to e^+ e^-}$[keV]        & 1.37 $\pm$ 0.05 & 1.24 & 
        2.06 \\
\end{tabular}
\end{table}

The amplitude
$\chi_P(x) = \langle 0 | T [ \Psi^1(\frac{1}{2} x)
\bar\Psi^2(-\frac{1}{2} x) ] | P \rangle$  is determined by the
Bethe-Salpeter equation, which in momentum space
reads~\cite{Res94,Mue94}:
\begin{equation}
        \chi_P(p) = S_1^F(p_1) \int \frac{d^4 p'}{(2\pi)^4}
                \left[ -i K(P,p,p') \chi_P(p') \right] S_2^F(p_2)\,.
\end{equation}
Here $p_{1/2}=\frac{1}{2}P+\!/\!-p$ denote the momenta of quark and
antiquark, $P$ is the four momentum of the bound state, $S^F$ is the
Feynman quark propagator and $K$ the irreducible quark interaction
kernel. Staying as close as possible to the non relativistic potential
model we make the following {\em Ansatz}: The propagators are assumed
to be of the free type, i.e. $S^F_i(p) = i/(p\!\!\!/ -
m_i+i\varepsilon)$, with an effective constituent quark mass $m_i$;
The kernel $K$ is assumed to depend only on the components of $p$ and
$p'$ perpendicular to P, i.e. $K(P,p,p') = V(p_\perp, p_\perp')$ with
$p_\perp := p - (pP/P^2)P$. Integrating in the bound state rest frame
over the time component $p^0$ and introducing the Salpeter (or
equal-time) amplitude $\Phi(p) = \int dp^0 \chi_P(p^0,\vec
p)|_{P=(M,\vec 0)}$ we obtain the Salpeter equation:
\begin{eqnarray}\label{eqn:Sal}
\Phi(\vec p) & = & \int \frac{d^3 p}{(2\pi)^3} \frac{
        \Lambda_1^-(\vec p) \gamma^0 [V(\vec p,\vec p')\Phi(\vec p')]
                          \gamma^0 \Lambda_2^+(-\vec p)}{M+m_1+m_2}
\nonumber \\
        & - & \int \frac{d^3 p}{(2\pi)^3} \frac{
        \Lambda_1^+(\vec p) \gamma^0 [V(\vec p,\vec p')\Phi(\vec p')]
                          \gamma^0 \Lambda_2^-(-\vec p)}{M-m_1-m_2},,
\end{eqnarray}
with the projectors $\Lambda_i^\pm(\vec p)=(\omega_i(\vec p)\pm
H_i(\vec p))/2\omega_i(\vec p)$, the Dirac Hamiltonian $H_i(\vec
p)=\gamma^0(\vec \gamma \cdot \vec p + m_i)$ and where $\omega_i(\vec
p) = \sqrt{m_i^2+\vec p^2}$.

\begin{figure}[hb]
  \centerline{\input{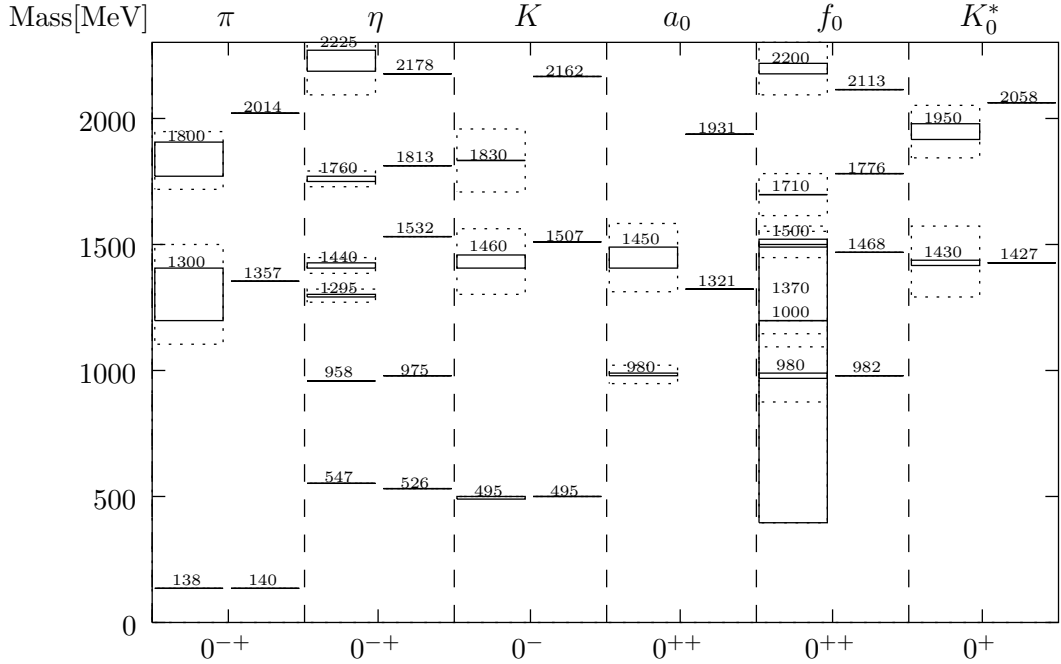}}
  \vspace{10pt}
  \caption{Comparison of the experimental (left side of each column, data
  from [2]) and calculated (right side of each column) spectrum for
 (pseudo)scalar mesons. The experimental uncertainty of the resonance
  position is
  indicated by a rectangular box, dashed boxes indicate the width.}
  \label{pseudo_scalar}
\end{figure}

The amplitudes $\Phi$ are calculated by solving the Salpeter
equation (\ref{eqn:Sal}) for a kernel containing a confining
interaction with a spin structure minimizing spin-orbit
effects:
\begin{equation}
         \int d^3 p' \left[ V(\vec p,\vec p') \Phi(\vec p') \right] = 
 -  \!\int \!d^3 p' v(\vec p-\vec p') \frac{1}{2} \left[ \gamma^0 \Phi(\vec p')
 \gamma^0 +  \Phi(\vec p') \right]\,.
\end{equation}
where $v$ in coordinate space is a linearly rising
potential $v(|\vec x_1-\vec x_2|) = a + b |\vec x_1-\vec x_2|$, and
 't Hoofts instanton induced interaction acting
exclusively on (pseudo)scalars:
\begin{equation}\label{eq:tHooft}
         \int \!d^3 p' \left[ W(\vec p,\vec p') \Phi(\vec p') \right] = 
 4 G^{(g,g')} \!\int\! d^3 p' w_\lambda(\vec p-\vec p') \left[ \gamma^5
 \mbox{tr} \left( \Phi(\vec p') \gamma^5 \right) +  \mbox{tr}
 \left(\Phi(\vec p')\right)  \right]\,.
\end{equation}
Here $G^{(g,g')}$ is a flavor matrix containing effective
coupling constants $g,g'$ and $w_\lambda$ is a regularizing Gaussian,
see~\cite{Mue94} for details. In order to calculate current matrix
elements in the Mandelstam-formalism~\cite{Mue95} we first construct
the meson-quark-antiquark vertex function $\Gamma_P(p) =
[S_1^F(p_1)]^{-1} \chi_P(p) S_2^F(p_2)$ in the rest frame from the
Salpeter amplitude by
\begin{equation}
         \Gamma(\vec p) := \Gamma_P(p_\perp)|_{P=(M,\vec 0)} = -i \int 
   \frac{d^4 p'}{(2\pi)^4} \left[ V(\vec p,\vec p') \Phi(\vec p') \right]
\end{equation} 
and then calculate the BS-amplitude for any on-shell momentum $P$ by a
boost $\Lambda_P$: $\chi_P(p) = S_{\Lambda_P} \chi_{(M,\vec
0)}(\Lambda_P^{-1}p) S_{\Lambda_P}^{-1}$.

The improvement of the results in Table~\ref{caption:table1} is
largely due to inclusion of the second term of equation
(\ref{eqn:Sal}), which is neglected in the non relativistic approach.
For other observables we refer to \cite{Mue94,Met96} concerning
spectra ,~\cite{Mue95,Gie96} for results on form factors, and
~\cite{Mue96} for an extensive discussion of $\gamma\gamma$-decays.
Here we will only briefly discuss the structure of (pseudo)scalar
mesons~\cite{Kle95,Rit96,Met96}. A comparison of their mass spectra
with experimental data is presented in Fig.\ref{pseudo_scalar}. It
shows, that the instanton induced interaction not only correctly
describes the splitting (and mixing) of the pseudoscalar ground state
nonet, but also leads to a particular structure for the scalar mesons:
It produces an almost pure flavor singlet state $f_0^1$ at roughly 1
GeV whereas the flavor octet states $f_0^8$, $a_0$, $K_0*$ are almost
degenerate at $\approx 1.4$ GeV.

This then leads to the following interpretation of experimental data
as the scalar $q\bar q$-flavor nonet~\cite{Kle95}: the
$f_0(1500)$ is not a glueball but the scalar (mainly)--octet meson.
The mainly--singlet state could correspond to the broad
$f_0(1000)$-state ($f_0(400-1200)$ of ~\cite{PDG96}), but there are
arguments that it is to be identified with the $f_0(980)$. The
isovector and isodoublet states then correspond to $a_0(1450)$ and
$K^*_0(1430)$, respectively.  The present model suggests that one
isoscalar and one isovector scalar state at 1 GeV is not of the
quarkonium type. Indeed, several calculations suggest that these
resonances are related to $K\bar K$-dynamics.  In this spirit, the
$f_0(1370)$ resonance is interpreted as the high energy part of the
broad $f_0(1000)$. Furthermore, we could identify the $f_J(1710)$ with
the first radially excited scalar state, provided its spin is indeed
$0^+$. However, in particular the $f_0(1500)$ was argued to have
properties incompatible with a pure $q\bar{q} $ configuration and was
suggested to possess a large glue component,
mainly because of the suppression of the $K\bar{K}$ decay mode,
see~\cite{amsclo}. In ~\cite{Rit96,Met96} it is shown,
that instanton effects can lead to a selective violation of the
OZI-rule for decays of scalars into pseudoscalars and that the
observed branching ratios can be explained in the framework of a
constituent quark model.

We conclude by citing some results on $\gamma\gamma$
decays~\cite{Mue96} of scalar and tensor mesons, which constitute a
sensitive test of the present approach, and thanking Eberhardt Klempt,
Claus M\"unz, Herbert Petry, J\"org Resag and Christian Ritter.

\begin{table}[!h]
        \caption{$\Gamma(M\to\gamma\gamma)[eV]$ of scalar and tensor
          mesons.}\label{caption:table2}
        \begin{tabular}{lcclcclcc}
        \multicolumn{1}{c}{M}& Exp.~\cite{PDG96}&
        \multicolumn{1}{c}{Calc.} & 
        \multicolumn{1}{c}{M}& Exp.~\cite{PDG96}&
        \multicolumn{1}{c}{Calc.} & 
        \multicolumn{1}{c}{M}& Exp.~\cite{PDG96}&
        \multicolumn{1}{c}{Calc.} \\
        \tableline
        $a_2(1320)$  & 1040 $\pm$ 90    & 734  & 
        $f_2(1270)$  & 2440 $\pm$ 300   & 2040 &
        $f_2'(1525)$ & 105  $\pm$ 17    & 121  \\
        $a_0(1450)$  &                  & 1390 &
        $f_0(980) $  & 560  $\pm$ 110   & 1750 &
        $f_0'(1500)$ & $<$ 170          & 161  \\
                     &                  &      &
        $f_0(1370)$  & 5400 $\pm$ 2300  &      &
                     &                  &      \\
\end{tabular}
\end{table}


\begin{references}
\bibitem{Bla90}
  Blask, W.H., Bohn, U., Huber, M.G., Metsch, B.C., Petry, H.R.,
  {\em Z. Phys.} {\bf A337}, 327 (1990).
\bibitem{PDG96}
  Barnett, R.M. {\em et al.}, (Particle Data Group), 
  {\em Phys. Rev.} {\bf D54}, 1 (1996) and
  1997 off-year partial update for the 1998 edition available on 
  the PDG WWW pages (URL: http://pdg.lbl.gov/). 
\bibitem{Res94} 
  Resag,  J., M\"unz, C.R., Metsch, B.C., Petry, H.R.,
  {\em Nucl. Phys.} {\bf A578}, 397  (1994).
\bibitem{Mue94}
  M\"unz, Claus R., Resag, J\"org, Metsch, Bernard C., Petry, Herbert R.,
  {\em Nucl. Phys.} {\bf A578}, 327 (1994).
\bibitem{Mue95} 
  M\"unz, Claus R., Resag, J\"org, Metsch, Bernard C., Petry, Herbert R., 
  {\em Phys. Rev.} {\bf C52}, 2110 (1995).
\bibitem{Gie96}
  Giersche, Wolfgang I.,M\"unz, Claus R.,
  {\em Phys. Rev.} {\bf C53}, 2554 (1996).
\bibitem{Mue96}
  M\"unz, C. R.,
  {\em Nucl. Phys.} {\bf A 609}, 364 (1996).
\bibitem{Kle95}
  Klempt, E., Metsch, B.C., M\"unz, C.R., Petry, H.R., 
  {\em Phys. Lett.} {\bf B361}, 160 (1995).
\bibitem{Rit96}
  Ritter, C., Metsch, B.C., M\"unz, C.R., Petry, H.R., 
  {\em Phys. Lett.} {\bf B380}, 431 (1996).
\bibitem{Met96}
  Metsch, B.C., Petry, H.R., {\em Acta Physica Polonica}, {\bf B27},
  3307 (1996).
\bibitem{amsclo} 
  Amsler, C., Close, F.,E., {\em Phys. Lett.} {\bf B353} 385 (1995);
  {\em Phys. Rev.}, {\bf D53} 295 (1996).  
\end{references}
\end{document}